\documentclass[12pt]{article}

\usepackage[body={17cm, 23cm, centered}]{geometry}
\usepackage{epsfig,verbatim,cite}

\usepackage{lmodern}
\usepackage[T1]{fontenc}
\usepackage[utf8]{inputenc}			

\usepackage{braket}

\usepackage{mathtext,marvosym,textcomp}
\usepackage{mathtools}
\usepackage{slashed}
\usepackage[usenames,dvipsnames,svgnames,table]{xcolor}
\usepackage{tikz}
\usepackage[ normalem]{ulem}
\usepackage{tocloft}
\usepackage{epsfig,amsmath,amsfonts,amssymb,arydshln}
\usepackage[makeroom]{cancel}
\usepackage{multirow}

 \usepackage[debug,pageanchor=false]{hyperref}
\hypersetup{colorlinks=true,linktocpage,breaklinks,
            urlcolor=blue,
            linkcolor=red,
            citecolor=blue
            }

\numberwithin{equation}{section}

\def\a{\alpha} 
\def\b{\beta} 
\def\g{\gamma} 
\def\d{\delta} 
\def\e{\epsilon}
 
\def\z{\zeta}

\def\l{\lambda} 
\def\m{\mu}

\def\p{\pi}

\def\q{\theta}
\def\s{\sigma} 
\def\t{\tau}  
\def\f{\phi}

\def\D{\Delta}

\def\L{\Lambda}

\def\fr{\frac}

\def\dt{\partial}

\def\ph{\phantom}

\def\mc{\mathcal}
\def\mA{\mathcal{A}}

\def\mF{\mathcal{F}}

\def\tz{\tilde{z}}

\def\tR{\tilde{R}}

\def\SS{\mathbb{S}}

\def\AdS{\mathrm{AdS}}
\def\HH{\mathrm{H}}

\def\rmO{\mathrm{O}}

\usepackage[pdftex]{pict2e}
\usepackage[dvipsnames]{xcolor}

\usepackage{tcolorbox}

\begin{document}

\begin{titlepage}
\ph{preprint}

\vfill

\begin{center}
   \baselineskip=16pt
   {\large \bf Uni-vector deformations, D0-bound states and DLCQ.  
   }
   \vskip 1cm
    Sergei Barakin$^{a,d}$\footnote{\tt barakin.serge@gmail.com },
    Kirill Gubarev$^{a,b,d}$\footnote{\tt kirill.gubarev@phystech.edu }, Edvard T. Musaev$^{c,a}$\footnote{\tt emusaev@theor.jinr.ru}
       \vskip .6cm
            \begin{small}
                          {\it
                          $^a$Moscow Institute of Physics and Technology, 
                          Laboratory of High Energy Physics, \\
                          9, Institutskii pereulok, 141702, Dolgoprudny, Russia
                           \\ 
                          $^b$Institute for Information Transmission Problems, 127051, Moscow, Russia\\
                          $^c$Bogoliubov Laboratory of Theoretical Physics, Joint Institute for Nuclear Research, \\ 6, Joliot Curie, 141980 Dubna, Russia\\
                          $^d$Institute of Theoretical and Mathematical Physics, Moscow State University, 119991, Russia} \\ 
\end{small}
\end{center}

\vfill 
\begin{center} 
\textbf{Abstract}
\end{center} 
\begin{quote}
We investigate uni-vector deformation in the Type IIA setup and show that the D0-brane background is mapped into itself (sedimentation), and other extremal backgrounds get bound with a dissolved D0-brane charge. Explicitly we generate F1-D0 and D2-D0 bound states background from uni-vector deformations. For the former we show that deformation of the non-extremal string gives the correct thermal F1-D0 bound state. We discuss relations between critical uni-vector deformations and DLCQ of M-theory.
\end{quote} 

\vfill
\setcounter{footnote}{0}
\end{titlepage}

\tableofcontents

\setcounter{page}{2}

\section{Introduction}

String theory (speaking more generally --- M-theory \cite{Townsend:1996xj}) is known to possess various limits in its moduli space that take it into regimes where dynamics becomes significantly different and/or gets simplified. One example of such a behaviour is provided by the open string on a background with a non-vanishing component of the Kalb--Ramond field in time-like direction, say  $B_{01}$. For the flat space-time there exists a critical value $B_{01} = (2\p \a')^{-1}$ when dynamics of the open string becomes non-relativistic \cite{Seiberg:2000ms,Seiberg:2000gc}. Moreover, it appears that this is not an exclusive feature of the open string dynamics, but the open string sector of what is called non-relativistic string theory (NRST). In this regime dynamics of the closed string becomes non-relativistic as well \cite{Gomis:2000bd}, D-brane states are represented by Dp-F1 bound states, where the fundamental string charge is dissolved along the Dp-brane world-volume \cite{Harmark:2000wv}.

Another example is provided by the BFSS matrix model and its relation to the discrete light-cone quantization (DLCQ) of the M2-brane \cite{Banks:1996vh}. The idea is based on the old idea advocated in \cite{Kogut:1972di,Weinberg:1966jm} where a system is observed in the infinite momentum frame, in other words, it is moving with respect to an observer with a velocity close to the speed of light. In this case dynamics of the system can be understood in terms of that of a set of point-like excitations, partons. In the infinite momentum frame quantization, that is basically the light-cone quantization, dynamics becomes non-relativistic in the sense that the dispersion relation for the energy $E=p_-$ takes the non-relativistic form with $p_+$ standing for the mass in denominator. For M-theory such partons are supposed to be D0-branes \cite{Townsend:1995af}.

The non-relativistic regime of string theory has close relation to non-commutative Yang--Mills theories on D-branes. Indeed, in both cases one considers the following transformation of the background geometry
\begin{equation}
\label{eq:openclosed}
    (g_\q + b_\q)^{-1} = (g+b)^{-1} + \q,
\end{equation}
where $\q = 1/2\, \q^{mn}\dt_m\wedge \dt_n$ is a bi-vector, $g$, $b$ denote some given background, $g_\q$, $b_\q$ denote the background that depends on $g$, $b$ and $\q$. If $b=0$ the relation \eqref{eq:openclosed} realizes the open-closed string map of \cite{Seiberg:1999vs} where a closed string metric $g_\q$ with a non-vanishing Kalb--Ramond field $b_\q$ is interpreted as the open string metric $g$ and the non-commutativity parameter $\q$. If $\q=\mathrm{const}$ and has no legs along the time direction, dynamics of the open string ends is described by non-commutative Yang--Mills theory. In this case it is possible to take the low energy limit $\a'\to 0$ to turn to a field theory. If $\q$ has components along time, say $\q^{01}\neq 0$, taking the limit is no longer possible, and the interpretation of the relation \eqref{eq:openclosed} changes. Now it is understood as literally a deformation of the initial closed string background $g$, $b$ by a bi-vector $\q$. In the special case $\q^{01} = \mathrm{const}$ and flat background space-time it is possible to take a different limit, where dynamics remains stringy, while the dispersion relation becomes non-relativistic \cite{Gomis:2008jt}.

The map \eqref{eq:openclosed} is a solution generating transformation, that is if $g$ and $b$ together with a dilaton field $\f$ solve supergravity equations, then $g_\q$, $b_\q$ and the new dilaton $\f_\q = \f - 1/4 \log \det g + 1/4 \log \det g_\q$ also solve the supergravity equation given certain conditions. The conditions are the following. First, the bi-vector is taken in the bi-Killing ansatz $\q^{mn} = r^{\a\b}k_\a{}^mk_\b{}^n$, where $k_\a{}^m$ is a Killing vector of the initial solution, and $r^{ab}=-r^{ba}$ is a constant matrix. Second, the r-matrix $r^{ab}$ satisfies the classical Yang--Baxter equation and the unimodularity constraint. The formalism of such bi-vector Yang--Baxter deformations has been developed in the series of works \cite{Klimcik:2002zj,Delduc:2013fga,Borsato:2016ose,Bakhmatov:2018apn,Bakhmatov:2018bvp}. A generalization to a wider class of deformations generated by poly-vector, which mix NS-NS and R-R fields, has been developed in \cite{Bakhmatov:2019dow,Bakhmatov:2020kul,Gubarev:2020ydf,Barakin:2024rnz,Gubarev:2024tks,Gubarev:2025qox}.

In the recent paper \cite{Barakin:2025jwp} behaviour of the extremal $D=10$ and $D=11$ brane backgrounds under poly-vector deformations has been investigated and a close relation to NRST and the open membrane theory (OM) has been observed. In particular, the Dp-F1 bound state background are generated by a bi-vector deformation $\q = \dt_0\wedge \dt_1$ from pure Dp-brane backgrounds. The same is true for M5-M2 bound states generated by tri-vector deformations generated by $\dt_0\wedge \dt_1 \wedge \dt_2$ in 11 dimensions. Moreover, such deformation translate respectively the F1-string and M2-brane backgrounds into themselves but with a different core charge. The same was true for 4-vector deformations applied to the D3-brane background. Such shifts in the space of solutions to D=10 Type II and D=11 supergravities were called sedimentation in analogy with the physical process.

In this paper we apply the same ideas to uni-vector deformations constructed in \cite{Gubarev:2025hvr} recently. These are generated by a single vector $\a = \a^m \dt_m$, that should be a Killing vector of the initial background for the deformation to map a solution to a solution. The deformation corresponds to motion in the space of solutions to Einstein--Maxwell dilaton (EMd) equations. The paper is structured as follows. In Section \ref{sec:sedim} we search for a solution that translates into itself under such a deformation, that appears to be an extremal point-like object (the D0-brane for the $D=10$ case and the extremal black hole in $D=4$ $\mc{N}=2$ supergravity). In Section \ref{sec:bound} we apply uni-vector deformations to F1 and D2-brane backgrounds to generate F1-D0 and D2-D0 bound states respectively. We show that deformations are able to generate the thermal F1-D0 bound state. In this Section we also investigate critical limit of uni-vector deformations, show that it is equivalent to an infinite boost, i.e. turning to the infinite momentum frame. Based on this observation we discuss relations between uni-vector deformations and BFSS/BMN matrix models in Conclusions.

\section{ Uni-vector deformations in Type IIA}
\label{sec:sedim}

The formalism of uni-vector deformations was developed in \cite{Gubarev:2025hvr} as a solution-generating method for Einstein--Maxwell-dilaton gravity. The key observation is that the lower-dimensional theory is obtained by Kaluza--Klein reduction of pure gravity in one dimension higher, so that the deformation is naturally formulated in terms similar to that of poly-vector deformations. For the latter the relation between the standard D=11 or D=10 Type II supergravity and the parent theory is the standard KK reduction ansatz. The generalized metric of double field theory (or exceptional field theory in the D=11 case) containing both the D=10 metric and the Kalb--Ramond field is naturally of the KK form. Poly-vector deformation are then given by exponentiation of certain generators of the full duality group $\rmO(10,10)$ (or groups in the exceptional series in D=11) branched under its GL(10) subgroup. For uni-vectors geometry behind deformations becomes much more transparent. The parent theory is the pure gravity theory with dynamics ruled by the standard Einstein--Hilbert action (probably with the cosmological term). The KK reduction ansatz for the $D$-dimensional metric is also standard
\begin{equation}
\label{eq:highermetric}
 ds^2 = e^{ 2\g (\f-\f_0)} g_{mn} dx^m dx^n + e^{2\b(\f-\f_0)}\big(dz + e^{\f_0}A_m dx^m\big)^2,
\end{equation}
with the constants $\b$ and $\g$ given by
\begin{equation}
    \gamma = \frac{1}{\sqrt{2(D-1)(D-2)}}, \quad \beta = - \sqrt{\frac{(D-2)}{2(D-1)}}
\end{equation}
to ensure that the $D-1$-dimensional theory is in the Einstein frame. In this language uni-vector deformations act on the lower-dimensional fields by a non-linear transformation,
\begin{equation}
\label{eq:field_transf}
\begin{aligned}
e^{2\beta \tilde \phi}& =\pm e^{2\gamma\phi}\alpha^k\alpha_k+e^{2\beta\phi}(1+A_k\alpha^k)^2 , \\
\tilde A_m&=\pm e^{-2\beta\tilde\phi}\Bigl(e^{2\gamma\phi}\alpha_m\pm e^{2\beta\phi}A_m(1+A_k\alpha^k)\Bigr), \\
\tilde g_{mn}&=e^{-2\gamma\tilde\phi}\bigl(e^{2\gamma\phi}g_{mn}\pm e^{2\beta\phi}A_mA_n\bigr)\mp e^{2(\beta-\gamma)\tilde\phi}\tilde A_m\tilde A_n .
\end{aligned}
\end{equation}
The deformation preserves the equations of motion provided the generating vector $\a = \a^{m}(x)\dt_m$ is a symmetry of the seed background,
\begin{equation}
\mathcal{L}_{\alpha}\phi=0,
\qquad
\mathcal{L}_{\alpha}g_{mn}=0 \, ,
\qquad
\mathcal{L}_{\alpha}A_m=0.
\end{equation}

A result of \cite{Gubarev:2025hvr} important for the present discussion is that the transformation \eqref{eq:field_transf}, apparently non-trivial, in the parent theory is simply a coordinate transformation. More precisely, the deformed configuration is obtained by
\begin{equation}
x'^M=e^{\xi}x^M ,
\end{equation}
with $\xi=z\,\alpha^m\partial_m $. Therefore, uni-vector deformations are geometrically trivial upstairs, while producing non-trivial Einstein--Maxwell--dilaton backgrounds after reduction. This also was demonstrated in \cite{Gubarev:2025hvr} by examples, where various scalars and tensors vanishing prior the deformation became non-zero after it. 

The equivalence between (general) uni-vector deformations and coordinate transformations in the parent theory allows to extend these ideas to constructions with parent theories other than the pure gravity. Moreover, it appear to be necessary to provide an interpretation of uni-vector deformations in terms of bound states similar to \cite{Barakin:2025jwp}. We will mainly consider the case where the parent theory is D=11 supergravity, when uni-vector deformations are naturally associated to D0-branes.

\subsection{Sedimentation of the pp-wave}

The pattern of correspondence between $p+1$-vector deformations and $p$-dimensional 1/2BPS objects observed in \cite{Barakin:2025jwp} suggests that the background that transforms into itself under uni-vector deformations must be that of a point-like object. Taking as the parent theory $D=11$ supergravity we expect this object to be the D0-brane of Type IIA theory. That is, transformations \eqref{eq:field_transf} of the fields of Type IIA theory should map the D0-brane background into itself under uni-vector deformations generated by $\a \dt_t$ with $\a$=const. Such vector is a direct analogue of poly-vectors $\g\, \dt_0\wedge\dots\wedge \dt_{p}$ with constant $\g$ used in \cite{Barakin:2025jwp}. To override the seeming obstruction that uni-vector deformations in the formulation of \cite{Gubarev:2025hvr} do not include transformations of Type IIA R-R fields, or equivalently of the 3-form in $D=11$, we use the fact that uni-vector deformations are equivalent to certain coordinate transformations in the parent theory. In the case of constant $\a$ the coordinate transformations are simply shear transformations
\begin{equation}
\label{eq:shear}
    t \to t + \a z, \quad z \to z,
\end{equation}
which we will often accompany with additional rescaling of the KK coordinate $z$. Since the D0-brane of Type IIA descends from the pp-wave of the D=11 theory, the question can be naturally replaced by the question of how pp-wave transforms under such diffeomorphisms.

It worth to note, that although it is obvious that boosts add momentum to a system, thus mapping pp-wave into itself, it is not obvious that such shear transformations do the same. While, as we show in Section \ref{sec:bound}, in the null limit these transformations are indeed equivalent, for finite $\a$ one should proceed carefully. Therefore, as in the case of all other branes for bi- and tri-vector deformation investigated in \cite{Barakin:2025jwp}, for the deformation to map such a solution to itself with a different charge one should start with a particular gauge choice. Therefore, the background we start with is the pp-wave in $D+1$ dimensions: 
\begin{equation}
    ds_{D+1}^2 = 2 dz dt + H dz^2 + ds_{\perp}^2.
\end{equation}
Note, that here $H = 1 + P/r^{D-3}$, that is $(t,z)$ are not the standard light-cone coordinates at $r\to \infty$, though related to them. Using the KK ansatz \eqref{eq:highermetric} we obtain the following $D$-dimensional solution
\begin{equation}
    \begin{aligned}
        ds_D^2 & = - H^{-\fr\g\b-1}dt^2 + H^{-\fr\g\b}ds_\perp^2, \\
        A & = e^{-\f_0}H^{-1}dt, \\
        e^{2\phi} & = H^{\fr1\b} e^{2\f_0}.
    \end{aligned}
\end{equation}
For $D=10$ we have the critical D0-brane background written in the Einstein frame. This solution is written in the gauge for which the potential $A$ does not vanish at $r\to \infty$. To turn to the standard gauge, where $A = e^{-\f_0}(H^{-1}-1)dt$ one performs gauge transformation, that is equivalent to $z \to z+t$ in the parent theory. It is worth to mention, that combination of such a transformation and the shear transformation is still not equivalent to a boost.

Perform now the uni-vector deformation as a coordinate transformation $t \to t - \a z$ in the parent theory accompanied by a rescaling  $z=  \m \tilde z$. The deformed D0-brane background is
\begin{equation}
    \begin{aligned}
        ds_D^2 & = - \m^{-\fr{2\g}{\b}}\big(H-2 \a\big)^{-\fr\g\b-1}dt^2 + \m^{-\fr{2\g}{\b}}\big(H-2 \a\big)^{-\fr\g\b}ds_{\perp}^2,\\
        \tilde{A}&=\m^{-1} e^{-\f_0}(H-2\a)^{-1}dt, \\
        e^{2\tilde{\phi}} & =\m^2 \big(H-2 \a\big)^{-\fr1\b}e^{2\phi_0},
    \end{aligned}
\end{equation}
where tilded variables denote deformed fields. For this to reproduce the same {\color{magenta}D}0-brane background as before with a different harmonic function $\tilde{H}$ the following equations must be satisfied
\begin{equation}
\label{eq:syseq}
    \begin{aligned}
        \m^{-\fr{2\g}{\b}}\big(H-2 \a\big)^{-\fr\g\b-1} & = \tilde{H}^{-\fr\g\b-1}\L^2, \\
        \m^{-1} e^{-\f_0}(H-2\a)^{-1} & = e^{-\tilde{\f}_0}\tilde{H}^{-1}\L,\\
        \m^2 \big(H-2 \a\big)^{-\fr1\b}e^{2\phi_0} & = \tilde{H}^{-\fr1\b}e^{2\b\tilde\f_0},\\
        \m^{-\fr{2\g}{\b}}\big(H-2 \a\big)^{-\fr\g\b} & = \tilde{H}^{-\fr\g\b}.
    \end{aligned}
\end{equation}
Here we assume, that the time direction is rescaled according to $t=\L^{-1}\tilde{t} $, which appears to be necessary. Multiplying the second and the third equations we obtain
\begin{equation}
    \m \, e^{(2\b-1)\f_0} = \L \, e^{(2\b-1)\tilde\f_0}.
\end{equation}
Taking the first equation to the power $\b$ and multiplying by the second equation in the power $-(\b+\g)$ we obtain $\L = \m$. This implies $\f_0 = \tilde\f_0$, that is natural as what we do is simply adding a momentum in 11 dimensions. Now, the first equation gives $\L^2(H-2\a) = \tilde H$, and setting $\L^{-2} = 1-2\a$ we get
\begin{equation}
    \tilde{H} = 1 + \fr{1}{1-2\a}\fr{P}{r^{D-3}}.
\end{equation}
Therefore the rescaling of time is needed to preserve the same asymptotics of the harmonic function at transverse infinity. Finally, we observe that the fourth equation of \eqref{eq:syseq} holds automatically. As the result we obtain
\begin{equation}
    \begin{aligned}
        ds_D^2 & = - \tilde{H}^{-\fr\g\b-1}d\tilde{t}^2 + \tilde{H}^{-\fr\g\b}d\tilde{s}_\perp^2, \\
        \tilde{A} & = e^{-\f_0} \tilde{H}^{-1}dt, \\
        e^{2\phi} & = \tilde{H}^{\fr1\b}e^{2\tilde\f_0},\\
        \tilde{H}& = 1 + \fr{1}{1-2\a}\fr{P}{r^{D-3}}.
    \end{aligned}
\end{equation}
As for the poly-vectors case we observe that increasing $\a$ corresponds to adding more {\color{magenta}D}0-brane charge to the background, or in the $D+1$-dimensional language to add momentum to the system. Taking $D=4$ and substituting the corresponding values for $\g$, $\b$, one obtains the extremal black hole solution to $D=4$ $\mc{N}=2$ supergravity. 

The critical value $\a =1/2$ corresponds to adding an infinite number of {\color{magenta}D}0-branes, equivalently, to turning to the infinite momentum frame. As we discuss in Section \ref{sec:bound}, in this limit the transformation becomes equivalent to boosting with infinite parameter. This equivalence allows to speculate on relations between critical uni-vector deformations in D=10 and DLCQ of M-theory (see the same Section \ref{sec:bound}).

\subsection{Particle on a uni-vector deformed background}

It is suggestive to investigate how probe charged massive objects behave on deformed background. As it is expected based on the form of the shear coordinate transformation, we will see that the probe charge coupled to the deformed background is shifted with respect to the initial charge by energy of the particle, while the energy does not change itself. This observation suggests that states with equal (proportional) energy and charges are natural for the deformation to map into themselves, give the appropriate rescaling of time (the rescale {\color{magenta}the (of?)} energy). Therefore, we start with the usual reparametrization-invariant particle action in $D+1$ dimensions $S_{D+1}=-m\int ds$. It is convenient to use the equivalent einbein form
\begin{equation}
S_{D+1}=\frac12\int d\tau\left[
e^{-1}G_{MN}\dot X^M\dot X^N - e\,m^2
\right].
\end{equation}
Substituting the KK metric gives
\begin{equation}
L
=
\frac12 e^{-1}
\left[
e^{2\g\f}g_{mn}\dot x^m\dot x^n
+ e^{2 \b\f}\big(\dot z+A_m\dot x^m\big)^2
\right]
-\frac12 e\,m^2 .
\end{equation}
Since the metric does not depend on $z$, the coordinate $z$ is cyclic, so its conjugate momentum is conserved:
\begin{equation}
p_z \equiv \frac{\partial L}{\partial \dot z}
=
e^{-1}e^{2\b\f}\big(\dot z+A_m\dot x^m\big).
\end{equation}
Call this constant $p_z = q$. In the $D$-dimensional theory this conserved KK momentum is interpreted as the electric charge under the graviphoton. Thus
\begin{equation}
\dot z+A_m\dot x^m = e\,q\,e^{-2\b\f}.
\end{equation}

To eliminate the cyclic variable $z$ while keeping the conserved charge $q$ fixed, we use a hybrid formulation of mechanics featuring both Hamiltonian and Lagrangian approaches. This is usually referred to as the Routhian mechanics, and is based on the following function
 \begin{equation}
L_R = L - p_z \dot z = L - q\dot z.
\end{equation}
The idea is to take into account the constraint $p_z=q= \text{const}$ in the standard Hamiltonian formulation and then go back to the Lagrangian. Since other canonical variables are not affected by the corresponding Lagrange transformation, it is sufficient to keep track only of $z$ and its momentum $q$.  Substitute $\dot z = e q e^{-2\b\f} - A_m \dot x^m$. Then
\begin{equation}
L_R
=
\frac12 e^{-1} e^{2\g\f}g_{mn}\dot x^m\dot x^n
-\frac12 e\left(m^2+e^{-2\b\f}q^2\right)
+ q\,A_m \dot x^m .
\end{equation}
This is the $D$-dimensional world-line action of a particle coupled to the D-dimensional metric $g_{mn}$, the scalar $\phi$, and the gauge field $A_m$. The interaction term is
\begin{equation}
S_{\text{int}} = q\int d\tau\, A_m(x)\dot x^m
= q\int_{\gamma} A ,
\end{equation}
where $\gamma$ is the particle world-line in $D$ dimensions and it is exactly of the Wess--Zumino type.

Eliminating the einbein $e$ from $L_R$ it is possible to go back to the Nambu--Goto form of the action
\begin{equation}
S_D
=
-\int d\tau\,
m_{\rm eff}(x)e^{\g\f}\sqrt{g_{mn}\dot x^m\dot x^n}
+ q\int_\gamma A.
\end{equation}
The ``mass'' however is not constant and depends on the dilaton field
\begin{equation}
m_{\rm eff}^2(x)=m^2+q^2 e^{-2\b\f}.
\end{equation}

Let us now perform uni-vector deformation of the background, for that we denote $X^M = (t,x^i,z)$, $i=1,2,D-1$ and assume that the metric does not depend on $t$ and $z$ coordinates. The uni-vector deformation in question is equivalent to the shear coordinate transformation $t \to  t+\a z$. From the $D+1$-dimensional point of view this is a diffeomorphism that mixes the compact direction with time and therefore reducing on the new $z$ is not the same as reducing on the original $z$. Instead it is a reduction along the Killing vector
\begin{equation}
K=\partial_{\tilde z}=\partial_z-\alpha\,\partial_t.
\end{equation}
Since the KK coordinate gets mixed with the time direction, at the level of the particle action the effect is in the change of charge, that gets mixed with the conserved energy. Indeed, the $(D+1)$-dimensional) action in the new coordinates is
\begin{equation}
S_{D+1}=\frac12\int d\tau\Big[
e^{-1}\big(e^{2\g\tilde{\f}}\tilde g_{mn}\dot{\tilde x}^m\dot{\tilde x}^n
+e^{2\b\tilde\phi}(\dot{\tilde z}+\tilde A_\m\dot{\tilde x}^\m)^2\big)
-e\,m^2
\Big].
\end{equation}
Now, the conserved charge is given by 
\begin{equation}
\tilde q \equiv p_{\tilde z}
=
e^{-1}e^{2\b\tilde\phi}(\dot{\tilde z}+\tilde A_\mu\dot{\tilde x}^\mu),
\end{equation}
that is the same expression but in terms of the new fields. Therefore, using the same procedure we arrive at the same particle action living on a different background with, however, a different charge
\begin{equation}
\tilde S_D
=
\int d\tau\left[
\frac12 e^{-1}e^{2\g\tilde\phi}\tilde g_{mn}\dot{\tilde x}^m\dot{\tilde x}^n
-\frac12 e\left(m^2+\tilde q^2e^{-2\b\tilde\phi}\right)
+\tilde q\,\tilde A_\mu\dot{\tilde x}^\mu
\right].
\end{equation}

To see the change in the charge consider transformation of the canonical 1-form
\begin{equation}
p_t\,dt+p_z\,dz
=
p_{\tilde t}\,d\tilde t+p_{\tilde z}\,d\tilde z.
\end{equation}
This implies $p_{\tilde t}=p_t$, $p_{\tilde z}=p_z-\alpha p_t$. If we define the conserved D-dimensional energy by
$E\equiv -p_t=-p_{\tilde t}$, and the original KK charge by
$q\equiv p_z$, then the charge in the sheared reduction is
\begin{equation}
\tilde q = p_{\tilde z}= q+\alpha E.
\end{equation}

So a state that was neutral in the first reduction, $q=0$, becomes charged in the sheared reduction $\tilde q=\alpha E $. We also see, why static BPS objects map into themselves under such a transformation: for them the energy is simply the mass, that is equal to charge. This is the physical reason behind the sedimentation procedure from the point of view of a probe charge. The same logic does not quite work for non-extremal objects that have mass and charge not proportional to each other. However, as we will see below uni-vector deformations of a non-extremal string reproduce correctly the thermal F1-D0-brane background. This however requires additional rescaling of the initial M2-brane charge, that might be related to the non-extremal energy/charge relation. To keep the text more sound we send a full discussion of these issues beyond the scope of the paper.

\subsection{The black-brane-like solution in D=4}

In \cite{Barakin:2025jwp} a note was made that for a poly-vector deformation to generate bound states or to change the core charge (to sediment the deformation), one has to have a seed harmonic function. In this section we show that the story is deeper than that in the sense that not any solution with a seed harmonic function suffices. For that we consider a black-brane like solution to Einstein--Maxwell-dilaton theory with cosmological constant ($\L$EMd) and show that upon a uni-vector deformation it changes locally but the asymptotic charge does not change. 

The black-brane-like background in question is a solution to equations following from the action 
\begin{equation}\label{GL5Action}
    S = \int d^4x \sqrt{- g} \Big( R[g_{mn}] - \fr12 \dt_m \f \, \dt^m \f - \fr14 e^{- \sqrt{3} \f} \mF^{m n} \mF_{m n} - 2 \L e^{\fr1{\sqrt{3}} \f}\Big),
\end{equation}
whose parent theory is the pure gravity with cosmological term. The corresponding field equations read 
\begin{equation}
\label{glDequations1}
\begin{aligned}
    R_{m n} - \frac{1}{2} R g_{m n} + \L e^{\frac{\sqrt{3}}{3}\phi} g_{m n} &= T_{mn}, \\ 
    \nabla_{m} ( e^{- \sqrt{3} \phi} \mF^{m n} ) &= 0\,, \\
    \Box \phi + \frac{\sqrt{3}}{4} e^{- \sqrt{3} \phi} \mF_{m n} \mF^{m n}  & = \frac{2\sqrt{3}}{3} \Lambda e^{\frac{\sqrt{3}}{3}\phi}\,, 
\end{aligned}
\end{equation}
where
\begin{equation}
    T_{mn}  =\frac{1}{2} \left( \dt_{m} \phi \, \dt_{n} \phi - \frac{1}{2} (\dt \phi)^2 g_{m n} \right)  + \frac{1}{2} e^{-\sqrt{3} \phi} \left( \mF_{m k} \mF_{n}{}^{k} - \frac{1}{4} \mF_{k l} \mF^{k l} g_{m n} \right).
\end{equation}
This theory appears to have the following vacuum solution
\begin{equation}\label{AdS2M2}
    \begin{aligned}
        \fr{d\tilde{s}^2}{R_{\AdS}^2} & = \fr{- dt^2 + dr^2}{r^2} + 2 \left( du^2 + \cosh^2 u \, dv^2 \right) , \\
        \fr{\mA}{R_{\AdS}} & = - \fr{dt}{r} \, e^{\fr{\sqrt{3}}2 \f_0}, \qquad \f = \f_0, \\
        \Lambda & = -\fr{3 \,  e^{-\fr1{\sqrt{3}} \f_0}}{4 \, R_\text{AdS}^2}.
    \end{aligned}
\end{equation}
The solution has geometry of $\AdS_2 \times \HH^2$, where $\HH^2$ is the two-dimensional hyperboloid. The Ricci scalar and the Maxwell field read
\begin{equation}
    R = - \frac{3}{R_\text{AdS}^2}, \quad \mF^{m n} \mF_{m n} = - \frac{2 e^{\sqrt{3} \f_0}}{R_\text{AdS}^2}.
\end{equation}

The form of the solution is similar to the near-horizon limit of an extremal $p$-brane solution. One would then expect that a uni-vector deformation changes its charge to recover a full $p$-brane solutions, which appears to be not the case. To see that we perform a uni-vector deformation of \eqref{AdS2M2} with the following  parameter
\begin{equation}
\label{univectroparameter}
    \alpha = \fr{Q \, e^{- \fr{\sqrt{3}}2 \f_0}}{2 R_{AdS}} \dt_t \,,
\end{equation}
where $Q=\text{const}$. The resulting deformed background is the following
\begin{equation}\label{AdS2M2def}
    \begin{aligned}
        \fr{d\tilde{s}^2}{R_{\AdS}^2} & = \fr{- H^{-\fr12}(r) dt^2 + H^{\fr12}(r) dr^2}{r^2} + 2 H^{\fr12}(r) \left( du^2 + \cosh^2 u \, dv^2 \right), \\
        \fr{\mA}{R_{\AdS}} & = - \fr{dt}{r} \, H^{-1}(r) \, e^{\fr{\sqrt{3}}2 \f_0} , \\
        \f & = \f_0 - \fr{\sqrt{3}}2 \ln H(r),
    \end{aligned}
\end{equation}
where $H(r) = 1 - \fr{Q}{r}$. The Ricci scalar and the Maxwell field change non-trivially
\begin{equation}
    \tR = - \fr{3 \left( 7 H^2(r) + 2 H(r) - 1 \right)}{8 R_\text{AdS}^2 H^{\fr52}(r)} \quad \text{and} \quad \tilde{\mF}^{m n} \tilde{\mF}_{m n} = - \frac{2 e^{\sqrt{3} \f_0}}{R_\text{AdS}^2 H^4(r)}.
\end{equation}

To calculate conserved charge we start with the conserved U(1) Noether current, that is 
\begin{equation}
    J^{n} = \nabla_{m}\left( e^{-\sqrt{3}\phi} \mF^{nm} \right), \quad \nabla_{n} J^{n} = 0 \quad \Rightarrow \quad \dt_{n}(\underbrace{\sqrt{-g} J^{n}}_{\mathcal{J}^{n}}) = 0 = \dt_{n} \mathcal{J}^{n}
\end{equation}
electric charge is
\begin{equation}
\begin{aligned}
    Q_{e} & = \int_{\Sigma} d^3 x  \sqrt{-g}J^{0} = \int_{\Sigma} d^3 x \sqrt{-g} \nabla_{m}\left(e^{-\sqrt{3}\phi} \mF^{0m}\right) = \int_{\Sigma} d^3 x \dt_{m}\left(\sqrt{-g} e^{-\sqrt{3}\phi} \mF^{0m}\right) = \\
    &= \int_{\dt\Sigma = \text{H}^2} d^2\s \sqrt{-g} e^{-\sqrt{3}\phi} \mF^{0r} \bigg|_{r\to \infty} = \text{Area}_{\text{H}^2} \left( - 2 R_\text{AdS}e^{- \frac{\sqrt{3}}{2} \phi_0} \right).
\end{aligned}
\end{equation}
For the harmonic function in question the solution should be interpreted as sourced by a point-like electric charge entering the equations as
\begin{equation}
    \nabla_{m}\left( e^{-\sqrt{3}\phi} \mF^{0m} \right) = \frac{Q_{e}}{\sqrt{-g}} \delta^3(\vec{x} - \vec{x}_0).
\end{equation}
This charge is however infinite and is proportional to area of the hyperboloid, which suggest that one should instead think of this as a smeared solution with a properly redefined harmonic function. On the other hand, in this case the solution is no longer asymptotically flat and requires regularization at infinity. To avoid these complications we simply forget about the infinities in $Q_e$ and treat it as a point-like charge, which does not affect our conclusions, that this charge does not depend on the deformation parameter. Therefore, although changing the background locally, the deformation does not add a charge asymptotically, implying that the seed solution has been chosen improperly.

It is straightforward to show that the magnetic charge of the deformed solution vanishes, and therefore the deformation does not add a magnetic charge either. For that one dualizes the action as
\begin{equation}
    \bar{\mF} = e^{-\sqrt{3}\phi} * \mF  \quad \Leftrightarrow \quad \bar{\mF}^{mn} = \frac{1}{2 \sqrt{-g}} e^{-\sqrt{3}\phi} \epsilon^{mnkl} \mF_{kl},
\end{equation}
and writes the conserved magnetic current as 
\begin{equation}
    \bar{J}^{n} = \nabla_{m}\left( e^{\sqrt{3}\phi} \bar{\mF}^{nm} \right).
\end{equation}
The magnetic charge is then
\begin{multline}
    Q_{m} = \int_{\Sigma} d^3 x \sqrt{-g}\bar{J}^{0} = \int_{\Sigma} d^3 x \sqrt{-g} \nabla_{m}\left(e^{\sqrt{3}\phi} \bar{\mF}^{0m}\right) = \int_{\Sigma} d^3 x \dt_{m}\left(\sqrt{-g} e^{\sqrt{3}\phi} \bar{\mF}^{0m}\right) = \\
    = \int_{\dt\Sigma = \text{H}^2} d^2\s \sqrt{-g} e^{\sqrt{3}\phi} \bar{\mF}^{0r} \bigg|_{r\to \infty} = \text{Area}_{\text{H}^2} \sqrt{-g} e^{\sqrt{3}\phi} \bar{\mF}^{0r} \bigg|_{r\to \infty} = \frac12 \text{Area}_{\text{H}^2} \mF_{uv} \bigg|_{r\to \infty} = 0.
\end{multline}
At this point it is not completely clear how deformation knows about the correct seed solution to choose, which one should change its asymptotical charge and which one should not. We believe, that such behaviour is quite interesting and requires further investigation.

\section{Bound states and limits }
\label{sec:bound}

In this section we extend the  property of poly-vector deformations advocated in \cite{Barakin:2025jwp} to generate bound states of certain branes with initial brane background. The standard bi-vector deformation were shown to generate dissolved F1-charge in certain setups, their S-dual naturally generate the D1-charge, quadri-vectors generate D3-brane charge and tri-vectors generate M2-brane charge. Given these results and similarity between uni-vector and poly-vector deformation, both practical and conceptual, we expect that the former should generate dissolved D0-brane charge, when applied to Type IIA backgrounds. We will consider two technically different cases: D2-D0 and F1-D0 bound states. The difference is in the standard ways of generating such bound state background from KK reduction of boosted M2-brane. In the first case the procedure is straightforward, while in the second case one has to start with the non-extremal membrane and take the extremal limit at the end.

\subsection{D2-D0 bound state}

For comparison of the extremal D2-D0 brane bound state to the solution obtained by a uni-vector deformation it is convenient to use the bound state background in the form similar to that used in \cite{Harmark:2000wv} and \cite{Barakin:2025jwp}. For that we start with  smeared extremal M2-brane extended along $(t,x^1,x^2)$, and let $z$ be the transverse direction. Defining
\begin{equation}
ds_7^2 \equiv dr^2+r^2 d\Omega_6^2,
\qquad
H_M=1+\frac{Q_{\text{M2}}}{r^6}.   
\end{equation}
we write the background as
\begin{equation}
\label{eq:M2_smeared}
\begin{aligned}
ds_{11}^2 & = H_M^{-\fr23}\left(-d\tilde{t}^2+dx^2+dy^2\right) +
H^{\fr13}\left(d\tz^2+ds_7^2\right), \\
A_3 &=(H_M^{-1}-\l)\,dt\wedge dx^1\wedge dx^2.
\end{aligned}
\end{equation}
In what follows we will prefer to use the gauge where the Type II gauge potential do not vanish at infinity, as this is what one reproduces from deformations most directly. To control the gauge we include $\l$=const in the 3-form potential.

Now perform a boost with rapidity $\beta$:
\begin{equation}
\begin{aligned}
t&=c\,\tilde{t}+s\,\tz, \\
z&=s\,\tilde{t}+c\,\tz,
\qquad
\end{aligned}
\end{equation}
where we use $c=\cosh\beta$, $s=\sinh\beta$ for compactness of expressions. 
Then the metric and the 3-form can be written in KK form as
\begin{equation}
\begin{aligned}
ds_{11}^2
=&-  H_M^{-\fr23}(c^2 - s^2 H_M) dt^2  + H_M^{-\fr23}(c^2 H_M - s^2)dz^2  \\
&+ 2 c \, s\, H_M^{-\fr23} (H_M- 1)dtdz + H_M^{-\fr23}(dx^2 + dy^2) + H_M^{\fr13}ds_7^2,\\
A_3  = & c (H_M^{-1}-\l)dt\wedge dx\wedge dy - s (H_M^{-1} - \l)dz\wedge dx\wedge dy.
\end{aligned}
\end{equation}
We now reduce the rescaled-boosted M2-brane background along $z$ using the  standard 11d/10d ansatz written in the string frame
\begin{equation}
\label{eq:KK11to10}
\begin{aligned}
    ds_{11}^2 & = e^{-\fr23(\f-\f_0)}ds_{10,\text{str}}^2+e^{\fr43(\f-\f_0)}(dz+g_s C_1)^2,\\
    A_3 & = g_s C_3 + B_2\wedge dz.
\end{aligned}    
\end{equation}
The resulting type IIA background in the string frame is as follows
\begin{equation}
\label{eq:D2-D0}
\begin{aligned}
ds_{10,\text{str}}^2
&=
H^{-\fr12}\left(-dt^2 + D (dx^2 + dy^2)\right) +
H^\fr12(dr^2+r^2d\Omega_6^2),\\
e^\f &= g_s\,D^{\fr12}H^{\fr14},\\
B_2 &= - s (D H^{-1} - \l)dx\wedge dy , \\
C_1 &= g_s^{-1} \fr{s}{c}(H^{-1}-1) dt,\\
C_3 &= g_s^{-1} c( DH^{-1}-\l)\,dt\wedge dx^1\wedge dx^2,
\end{aligned}
\end{equation}
where
\begin{equation}
    \begin{aligned}
        H & = \cosh^2 \b H_M - \sinh^2 \b = 1 + \fr{\cosh^2 \b Q_{\text{M2}}}{r^5},\\
        D^{-1} & = \fr{1}{\cosh^2 \b} + \fr{\sinh^2 \b}{\cosh^2 \b}H^{-1}.
    \end{aligned}
\end{equation}
It is straightforward to see that this background corresponds to the D2-brane carrying a D0-brane charge dissolve as magnetic flux. Indeed, the KK vector $C_1$ is the RR 1-form of type IIA, which couples electrically to a D0-brane, so the reduced object carries a D0  charge. The reduced background also has $B_{12}\neq 0$, i.e. a magnetic 2-form along the D2 spatial directions. In the D2 world-volume description, the gauge-invariant combination $\mathcal F = B_2 + 2\pi\alpha' F$ 
has a spatial component $\mathcal F_{12}$, and the D2 Wess--Zumino coupling
\begin{equation}
    \int_{\text{D2}} C_1\wedge \mathcal F    
\end{equation}
shows that magnetic flux on the D2 induces D0 charge. This is exactly the statement that the D0-branes are dissolved in the D2 as magnetic flux. At large $r$ in the gauge $\l=1$ we have
\begin{equation}
C_3 \simeq -g_s^{-1}\,\frac{Q_{\text{M2}}}{r^5}\,d t \wedge dx\wedge dy,
\qquad
C_1 \simeq -g_s^{-1} \frac{\sinh \b \cosh \b\, Q_{\text{M2}}}{r^5}\,d t.
\end{equation}
For charges this implies $Q_{\text{D2}} \propto Q_{\text{M2}}$, $Q_{\text{D0}} \propto \sinh \b \cosh \b Q_{\text{M2}}$, and therefore the boost parameter controls the ratio
\begin{equation}
\frac{Q_{\text{D0}}}{Q_{\text{D2}}} = \sinh \b \cosh \b.
\end{equation}

Let us now compare this to the result of a uni-vector deformation. Since uni-vector deformations $t \to t + \a z$ considered here are parabolic transformations, in contrast to boosts, that are hyperbolic transformations, one naively cannot expect to arrive at the same results here. It indeed be true, moreover after a straightforward deformation (even with rescaling of $z$) the resulting background has a slightly different form. This difference however has a nice interpretation given the nature of the deformation. 

We now start with the same smeared background \eqref{eq:M2_smeared}, fix the gauge $\l=1$ and perform a uni-vector deformation of the corresponding D2-brane background equivalent to the following shear transformation
\begin{equation}
    t\to t+\a z, \quad z \to \m z.
\end{equation}
As we will see the rescaling of $z$ is necessary to reproduce the correct behaviour of the final harmonic function. After the transformation the $D=11$ background becomes
\begin{equation}
    \begin{aligned}
        ds_{11}^2  = & - H_M^{-\fr23}dt^2  - 2 \a H_M^{-\fr23}dtdz + \left(\m^2 H_M^{\fr13} - \a^2 H_M^{-\fr23}\right)dz^2 \\
        & + H_M^{-\fr23}(dx^2 + dy^2) + H_M^{\fr13}ds_7^2,\\
        A_3  = &\ H_M^{-1}dt\wedge dx\wedge dy + \a H_M^{-1}dz\wedge dx\wedge dy.
    \end{aligned}
\end{equation}
After reduction we arrive at the following Type IIA background
\begin{equation}
    \begin{aligned}
        ds_{10}^2 & = - \m^2 H^{-\fr12} dt^2 + H_M^{-1}H^{\fr12}(dx^2 + dy^2) + H^{\fr12}ds_7^2, \\
        B_2 & = \a H_M^{-1}dx\wedge dy, \\
        e^{\f} & = g_s H_M^{-\fr12}H^{\fr34} ,\\
        C_1 & = - g_s^{-1} \a H^{-1} dt, \\
        C_3 & = g_s^{-1}H_M^{-1}dt\wedge dx\wedge dy,
    \end{aligned}
\end{equation}
where $H = \m^2 H_M - \a^2$. It is possible to match this solution to \eqref{eq:D2-D0} for all fields except $C_3$. This is done by rescaling $t \to \m^{-1} t $ and taking
\begin{equation}
    \m = \cosh \b, \quad \a = \sinh \b.
\end{equation}
The result is the following
\begin{equation}
\label{eq:D2-D0-prime}
\begin{aligned}
ds_{10,\text{str}}^2
&=
H^{-\fr12}\left(-dt^2 + D (dx^2 + dy^2)\right) +
H^\fr12(dr^2+r^2d\Omega_6^2),\\
e^\f &= g_s\,D^{\fr12}H^{\fr14},\\
B_2 &= - s D H^{-1} dx\wedge dy , \\
C_1 &= g_s^{-1} \fr{s}{c}(H^{-1}-1) dt,\\
C_3 &= g_s^{-1} \fr{1}{c}  DH^{-1}\,dt\wedge dx^1\wedge dx^2,
\end{aligned}
\end{equation}
with the same $D$ as before. As announced, the difference is in $C_3$, where $\cosh\b$ stands in the denominator, rather than in the numerator. This however is still a D2-D0 bound state with dissolved D0-charge. Turning to the gauge where $C_1$ and $C_3$ vanish at $r\to\infty$ we obtain the following relations for the R-R charges
\begin{equation}
    \begin{aligned}
        Q_{\text{D2}} \sim \cosh \b Q_{\text{M2}}, \quad Q_{\text{D0}} \sim \sinh\b \cosh \b Q_{\text{M2}}, \quad \fr{Q_{\text{D0}}}{Q_{\text{D2}}} = \sinh \b = \a.
    \end{aligned}
\end{equation}
Hence, what changes with respect to the standard procedure is the way how the effective D2-brane charge is related to the initial M2-brane charge. It depends on the deformation parameter in such a way, that ratio of the D0 and D2 charges vanishes for $\a \to 0$.

As the result we observe that uni-vector deformation of the D2-brane background Type IIA theory realized as shear coordinate transformation of the M2-brane in the parent D=11 theory is equivalent to adding a dissolved D0-brane charge. Hence, in this case two procedures: i) boost+KK-reduction and ii) shear+rescalings+reduction give physically the same background. It is possible to make the correspondence mathematically precise, however the corresponding calculations are quite messy. For this reason we send the calculational details into Appendix \ref{sec:D2-D0-precise} as a support of this statement.

\subsection{F1-D0 bound state}

The F1-D0 extremal bound state realized as KK reduction of the boosted M2-brane is more tricky. The reason is that in this case one has to boost along the M2-brane itself, that follows from the realization of the fundamental string as the M2-brane wrapping the KK cycle. Due to Lorentz invariance the boost gives again the M2. Technically, the reason is that one does not have function $H_{M2}$ standing in different powers as prefactors of $dt^2$ and $dz^2$. This suggest the standard trick, that is to start with the non-extremal brane:
\begin{equation}
\label{eq:M2-thermal}
\begin{aligned}
ds_{11}^2 & =  H^{-\fr 23}\left(-f\,dt^2+dx^2+dz^2\right)+H^{\fr 13}d\Sigma_8^2 ,\\
A_3 & = \coth\delta\,\left(H^{-1}-1\right) \,dt\wedge dx\wedge dz.
\end{aligned}
\end{equation}
The additional blackening function 
\begin{equation}
f(r)=1-\frac{r_0^6}{r^6},
\end{equation}
does the trick. The harmonic function is given by the usual expressions
\begin{equation}
    H(r)=1+\frac{r_0^6\sinh^2\delta}{r^6}.
\end{equation}
One then boosts along $z$, performs the KK reduction and takes the extremal limit $r_0 \to 0$, while $r_0^6\sinh^2\delta = Q_{\text{F1}}$ fixed. The result is
\begin{equation}
\begin{aligned}
ds_{10}^2 & = H_1^{-1}H_0^{-\fr12}\left(-dt^2\right) + H_1^{-1}H_0^{\fr12}dx^2 + H_0^{\fr12}dx_\perp^2 , \\
e^\phi& =g_s H_1^{-\fr12}H_0^{\fr34} , \\
B_2 & =\left(H_1^{-1}-1\right)\,dt\wedge dx , \\
C_1 & =g_s^{-1}\left(H_0^{-1}-1\right)\,dt ,
\end{aligned}
\end{equation}
where
\begin{equation}
    H_1 = 1 + \fr{Q_{\text{F1}}}{r^6}, \quad H_0 = 1 + \fr{Q_{\text{D0}}}{r^6}.
\end{equation}

We now show that one is able to reproduce precisely this answer using the uni-vector deformation procedure of the initial non-extremal background. Start with \eqref{eq:M2-thermal} and  perform the shear and rescaling transformations $t \to t+\alpha z$, $z \to \mu z$ to obtain
\begin{equation}
\begin{aligned}
ds_{11}^2 & =
H^{-\fr23}\Bigl(-fdt^2-2\alpha f\,dt\,dz+\Delta\,dz^2+dx^2\Bigr)
+H^{\fr13}d\Sigma_8^2 ,\\
A_3 & = \mu\,\coth\delta\,\left(H^{-1}-1\right) \,dt\wedge dx\wedge dz ,
\end{aligned}
\end{equation}
where $\Delta(r)=\mu^2-\alpha^2 f(r)$.  After reduction along $z$ using the standard Kaluza--Klein ansatz \eqref{eq:KK11to10} we get
\begin{equation}
\begin{aligned}
ds_{10,\mathrm{str}}^2 & =\Delta^{1/2}\left[
H^{-1}\left(-\frac{\mu^2 f}{\Delta}dt^2+dx^2\right)
+f^{-1}dr^2+r^2 d\Omega_7^2 \right] ,\\
e^\phi &= g_s H^{-\fr12}\Delta^{\fr34} ,\\
B_2 & =\mu\,\coth\delta\,\left(H^{-1}-1\right)\,dt\wedge dx ,\\ 
C_1 & =-g_s^{-1}\a f\D^{-1}\,dt .
\end{aligned}
\end{equation}

To put the result into the desired two-function form, we have to impose the normalization $\mu^2-\alpha^2=1$, taking $\mu=\cosh\beta$, $\alpha=\sinh\beta$. Then
\begin{equation}
\Delta(r)=\cosh^2\beta-\sinh^2\beta\,f(r)=1+\frac{r_0^6\sinh^2\beta}{r^6} \equiv H_0(r) ,
\end{equation}
is the harmonic function corresponding to the D0-brane charge. Rescaling  $t \to \cosh\beta^{-1}\, t $ we arrive at
\begin{equation}
\begin{aligned}
ds_{10,\mathrm{str}}^2 &  = H^{-1}H_0^{-\fr12}\left(-f\,d t^2\right) + H^{-1}H_0^{\fr12}dx^2 + H_0^{\fr12}\left(f^{-1}dr^2+r^2 d\Omega_7^2\right) , \\
e^\phi& = g_s H^{-\fr12}H_0^{\fr34}, \\
B_2 & =\coth\delta\,\left(H^{-1}-1\right)\,d t\wedge dx ,\\
C_1&=-g_s^{-1}\tanh\beta\,f H_0^{-1}\,d t .
\end{aligned}
\end{equation}
Up to a constant gauge shift in the RR 1-form potential, that in $D=11$ corresponds to shear transformation $z \to z + \l t$ with an appropriate $\l$=const, the background is the standard two-function non-extremal F1--D0 solution. Naturally, taking the extremal limit $r_0\to 0$, $\delta\to\infty$,
 $\beta\to\infty$, while keeping fixed
\begin{equation}
Q_{\mathrm{F1}}=r_0^6\sinh^2\delta,
\qquad
Q_{\mathrm{D0}}=r_0^6\sinh^2\beta ,
\end{equation}
one reproduces the extremal F1-D0  bound state background

It is worth to stop here and discuss the result obtained prior taking the extremal limit. The resulting background is a non-extremal solution to Type IIA supergravity equations, that has a regular Killing horizon and a single well-defined Hawking temperature
\begin{equation}
    T = \fr{6}{4 \p r_0 \cosh \d \cosh \b}.
\end{equation}
This is a thermal F1–D0 bound state generated by uni-vector deformation from thermal F1 state in Type IIA theory (see \cite{Cvetic:1995yq,Cvetic:1996gq,Tseytlin:1996bh}). Given the conceptual similarities between uni-vector and poly-vector deformations, this result seems to be in tension to the  discussion in \cite{Barakin:2025jwp}, where bi-vector deformation failed to produce known thermal bound state backgrounds. Since this is clearly not the case for uni-vector deformations, one becomes interested to return to the bi-vector case and reconsider poly-vector deformations of non-extremal branes. This stands beyond the scope of this paper, however we will briefly discuss this in the Conclusions section.

\subsection{Uni-vector deformations in the Sen--Seiberg limit}

It appears, that the correspondence between the standard procedure involving boosts and reductions and uni-vector deformations becomes more general in the null boost limit, that we turn to now. It appears possible to identify the boost transformation and the uni-vector deformation, that is a shear coordinate transformation, in the infinite boost limit. This is the limit used in the DLCQ of M-theory in \cite{Sen:1997we,Seiberg:1997ad} to recover the BFSS model \cite{Banks:1996vh}, therefore we refer to it as the Sen--Seiberg limit. For finite rapidity, reduction of the boosted smeared M2 is an ordinary space-like KK reduction and gives the standard D2-D0 bound state, while in the infinite-boost/DLCQ limit, the boosted KK circle becomes null. The idea is that in coordinates adapted to that almost-null circle, the exact boost degenerates into a shear.

Start from the exact boosted circle and rewrite it in a slightly different coordinate frame w.r.t. that in the previous section.  A boost with rapidity $\b$ from $(t,z)$ to $(T,Y)$ in the corresponding light-cone coordinates takes the following form
$$
x^+ = e^\beta X^+,\qquad x^- = e^{-\beta}X^-,
\qquad X^\pm=\frac{T\pm Y}{\sqrt2}.
$$
We will eventually tend $x^+$ to be a null direction, therefore the reduction coordinate $Y$ is 
$$
Y=\frac{1}{\sqrt2}\left(e^\beta x^+ - e^{-\beta}x^-\right).
$$
The corresponding Killing vector $
\partial_Y=e^\beta\,\partial_{x^+}-e^{-\beta}\,\partial_{x^-}$ is space-like for finite $\beta$. For $\beta\to\infty$, it becomes proportional to $\partial_{x^+}$, i.e. the circle becomes null.

To turn to the Sen--Seiberg null limit let the space-like M-theory circle has radius $R_s$. To get a finite null circle, take
$\beta\to\infty$, $R_s\to 0$ while $R_- = \frac{e^\beta}{\sqrt2}R_s
$ is held fixed. Then the identification along $Y$ becomes a null identification along $x^+$:
$$
\Delta x^+ = 2\pi R_-,
\qquad
\Delta x^- \sim e^{-2\beta}\to 0,
$$
that is the standard DLCQ/null-reduction limit. 

To show that the boost in this limit is equivalent to shear we choose coordinates adapted to the almost-null circle: $\zeta \equiv Y$, $\tau \equiv x^-$. Expressing $x^\pm$ back in terms of these adapted coordinates we get
$$
x^+ = \sqrt2\,e^{-\beta}\zeta + e^{-2\beta}\tau,
\qquad
x^-=\tau.
$$
Finally, rescaling the circle coordinate as $\tilde\zeta \equiv \sqrt2\,e^{-\beta}\zeta$ we obtain
$$
x^+ = \tilde\zeta + \epsilon\,\tau,
\qquad
x^-=\tau,
\qquad
\epsilon = e^{-2\beta},
$$
that in the limit $\e \to 0$ is exactly a shear
$$
x^+ \to x^+ + \epsilon x^-,
\qquad
\epsilon\to 0.
$$

So the shear picture is not equivalent to a finite boost, but it is the correct description of the null limit of the boost. This implies, that all bound states obtained by the boost-reduction procedure in the infinite boost frame can equivalently be obtained by the uni-vector deformation. In addition to the M2-p $\to$ D2-D0 reduction discussed above such an algorithm includes wM2-p $\to$ F1-D0, wM5-p $\to$ D4-D0 and M5-p $\to$ NS5-D0 reductions, where 'w' denotes wrapping of the membrane around the KK circle and 'p' stands for KK momentum.

The subtlety here is the following. While for the M2-p $\to$ D2-D0 equivalence between the boost-reduction and uni-vector deformations can be shown for arbitrary boost/deformation parameter, it seems not possible for other reductions. Although, at the moment the obstruction seems to be technical, and manifests itself at various numerical prefactors preventing two background to coincide, we believe there is a deeper reason for such a discrepancy. Indeed, in contrast to the the M2-p $\to$ D2-D0, the other reductions cannot be performed directly at the level of BPS objects. Instead, one has to start with a non-extremal M2 or M5 brane, boost it, reduce and then take a specific extremal limit. Otherwise, the procedure simply reproduces the initial brane charge. On the other hand, we observed in \cite{Barakin:2025jwp} that poly-vector deformations fail to reproduce thermal bound states, i.e. bound states of non-extremal branes. Although here we are dealing simply with coordinate transformations, and the resulting backgrounds are definitely solutions to Type IIA supergravity equations, the problem with reproducing the correct charges might have the same (still unclear) origin.

\subsection{DLCQ}

In this section we would like to highlight similarities between bi-vector deformations, understood as adding space-time non-commutativity parameter, and uni-vector deformations along a constant time-like Killing vector (analogous to shear coordinate transformations in the parent theory). For that let us briefly recall the relation between non-relativism in string theory and bi-vector deformations. This relation was discussed in \cite{Blair:2020ops} in terms of families of non-degenerate generalized metrics, interpolating between relativistic and non-relativistic Gomis--Ooguri backgrounds \cite{Gomis:2000bd}.

Let us start with a bi-vector deformation of the flat space-time background. Using the general prescription
\begin{equation}
    (g+2 \pi \a'  b)^{-1} = G^{-1} + \fr{1}{2\p \a'} \beta
\end{equation}
relating the initial metric $G$, the deformation bi-vector $\b$ and the deformed metric $g$ and Kalb--Ramond field $b$, we obtain
\begin{equation}
    \begin{aligned}
        ds^2 & = \fr{1}{1-\g^2}(-dt^2 + dx^2) + d\vec y^2,\\
        2 \p \a' b & = \fr{\g}{1-\g^2}dt\wedge dx.
    \end{aligned}
\end{equation}
The deformation bi-vector has been taken in the form $\b = 2\p\a'  \g \dt_t \wedge \dt_x$. To relate this to the non-relativistic limit of Gomis--Ooguri it is convenient to introduce a variable $\m^{-2} = 1-\g^2 $. The background then takes the form
\begin{equation}
    \begin{aligned}
        ds^2 & = \m^2(-dt^2 + dx^2) + d\vec y^2,\\
        2\p\a'b & =  \m^2 \sqrt{1 - \m^{-2}}dt\wedge dx.
    \end{aligned}
\end{equation}

Now taking the limit $\m\to \infty$ reproduces the non-relativistic limit of Gomis--Ooguri \cite{Gomis:2008jt}. In this limit the slope parameter $\a' = \e \a'_{eff}$ tends to zero as the regulating parameter $\e \to 0$ such that $\a'_{eff}$ is fixed.  Such rescaled  $\a'_{eff}$ non-commutativity plays the role of the slope parameter in the effective string theory, that appears to have non-relativistic spectrum. The same is true for open string theory in the presence of critical space-time non-commutativity parameter $\g \to 1$. Therefore, one may say, at least in this particular example of the flat space-time, that bi-vector deformations transform the background in such a way, that string dynamics ends ups to be that of NRST in the critical limit. Another observation supporting this interpretation is that backgrounds of Dp-F1 bound states playing the role of Dp-branes in NRST according  to \cite{Harmark:2000wv} are related to the usual Dp-brane backgrounds by certain bi-vector deformations. The same is true for tri-vector deformations, that add M2-brane dissolved charge and thus in the critical limit are supposed to give what is called the OM theory \cite{Gopakumar:2000ep}.

We would like now to extend such an interpretation to uni-vector deformations. Based on the examples provided by the pp-wave sedimentation, the D2-D0 brane bound state background and the discussion in the previous section one may suggest a general interpretation that  uni-vector deformation adds momentum (in the Sen--Seiberg limit). The momentum is along the KK circle, that in the lower dimensional theory equivalent to adding D0-brane states.

The equivalence of between shear transformations and infinite boost can be shown in a different, more convenient form
Start with the $D+1$-dimensional Minkowski background
\begin{equation}
    ds^2 = -dt^2 +dz^2 + ds_\perp^2.
\end{equation}
and perform the coordinate transformation $t\to t+\a z$, that corresponds to a certain uni-vector deformation in the theory in $D$ dimensions. Adding the following rescaling suggested by the pp-wave sedimentation procedure
\begin{equation}
    \begin{aligned}
        t & \to (1-\a^2)^{\fr14}t,\\
        z & \to (1-\a^2)^{-\fr14}z,
    \end{aligned}
\end{equation}
we arrive at the following
\begin{equation}
    ds^2 = - (1-\a^2)^\fr12dt^2 - 2\a dtdz + (1-\a^2)^{\fr12}dz^2 + ds_\perp^2.
\end{equation}
In the limit $\a \to 1$ this turns into the interval written in light-cone coordinates $t=u$, $z=v$. Therefore, in the critical limit  $\a\to1$ uni-vector deformation of the flat space-time give the light-cone. 

If dynamics of M2-brane is considered on a flat space-time deformed by shear transformations, in the critical limit it becomes simply the M2-brane in light-cone coordinates. After the proper choice of gauge for the world-volume metric and regularizing functions of the embedding coordinates $X^\m(\s)$ by matrices one ends up with a matrix model, equivalent to the BFSS model \cite{Banks:1996vh}.  Its mass deformation, known as the BMN model \cite{Berenstein:2002jq}, describes M-theory on the maximally supersymmetric pp-wave background (see also \cite{Dasgupta:2002hx}). The latter differs from the pp-wave considered above by a non-vanishing 4-form flux and is known to be the Penrose limit of $\AdS_4\times \SS^7$ \cite{Penrose1976PlaneWaveLimit,Gueven:2000ru,Blau:2002dy}. It is worth to mention the differences between the Penrose limit and the infinite momentum frame limit (null boost or turning to the light-cone coordinates). The IMF limit selects states with very large momentum in a given theory, these are precisely the states one obtains in the light-cone quantization. In the Penrose limit one focuses at the region of space-time near null geodesics, that is restricts the full space-time geometry  to the region seen by objects moving along null trajectories. As a result we observe striking similarities between bi-vector deformations, that in the critical limit send the string into the NRST sector, and uni-vector deformations, that are naturally related to the DLCQ of M-theory, where it is described in terms of D0-branes.

\section{Conclusions and discussions}

In this paper we extend the logic of \cite{Barakin:2025jwp} to uni-vector deformations asking two questions: i) what is the natural object in string/M-theory corresponding to such deformations; and ii) what happens in the limit of a critical uni-vector deformation. In \cite{Barakin:2025jwp} it has been shown that for bi-vector deformations the natural object is the fundamental string and the critical limit corresponds to going to the non-relativistic phase of string theory (NRST). The first statement is supported by the fact that the fundamental string background under the abelian bi-vector deformation generated by $\b = \g \dt_0\wedge \dt_1$ maps into itself with a different core charge. We call this map sedimentation. The second statement is supported by the observation that such bi-vector deformation starting with the Dp-brane background  generate Dp-F1 bound states, that is the proper generalization of Dp-branes in NRST according to \cite{Harmark:2000wv}. Also action of the closed string on such deformed flat space-time reproduces the Gomis--Ooguri limit \cite{Gomis:2008jt}.

Here we translate this statements to uni-vector deformations and observe that the natural object is the pp-wave in the parent theory or the D0-brane in D=10 Type IIA theory. This is precisely the background for which uni-vector deformations act as sedimentation procedure. Naturally we check that applied to F1 and D2 backgrounds of Type IIA such deformations produce F1-D0 and D2-D0 bound state background respectively. More generally one finds that uni-vector deformations give sedimentation for point-like 1/2BPS states as expected. To demonstrate that this is no coincidence but rather a result of a peculiar interaction between deformations and initial background, we check uni-vector deformation for a black-brane-like background in $\Lambda$EMd theory. We find that although the background non-trivially changes locally, its asymptotically measured charge does not change. It is important to mention, that the fact that pp-wave is reproduced under shear coordinate transformations is not obvious in the sense, that it naturally is invariant under boost transformations. The latter are parabolic transformation, while the former are hyperbolic, and the two are equivalent only in the null limit (to be discussed below). However, for this particular setup after a proper rescaling they give the same, nicely integrating the standard boost transformations into the general picture of poly-vector deformations.

An interesting observation we made is that uni-vector deformations produce the correct thermal F1-D0 bound states when starting with the non-extremal string. This seems to be in tension with the discussion in \cite{Barakin:2025jwp} related to failure of bi-vector deformations of thermal backgrounds to produce known in the literature thermal bound state solutions. There we assumed, that the reason might be that poly-vector deformations in general add only BPS dissolved states, i.e. states at zero temperature. And since the procedure should be understood as performed for states at thermal equilibrium, one should look for a different class of deformations, that carry information about temperature of the states they add. On the other hand uni-vector deformations in their spirit do not differ from deformations generated by poly-vectors, that allows to expect that the same poly-vectors should produce correct thermal bound states when started with non-extremal branes. It therefore would be interesting to investigate this tension more closely and reconsider the construction of thermal bound states, both the standard and by poly-vector deformations, to eventually match them. A first natural step here would be to systematize the non-extremal sector and to extend the analysis to other thermal bound states, clarify when sedimentation survives away from extremality, and determine whether the relevant observables are best organized by horizon thermodynamics, blackfold variables, or dual matrix-model data.

To answer the question of what happens in the critical limit of uni-vector deformations, we consider the Sen--Seiberg limit of such deformations and find that in the parent theory they become equivalent to an infinite boost. We show explicitly that upon performing such a transformation on the flat space-time in the parent theory and properly rescaling the time and the KK cycle coordinates, one ends up in the light-cone frame. The latter is the same as the infinite momentum frame achieved by boosting the system to a speed close to the speed of light. As it has been discussed in a series of works \cite{Banks:1996vh,Seiberg:1997ad,Sen:1997we,Berenstein:2002jq} M-theory in this limit can be described in terms of a matrix model relevant to a description of a set of D0-branes (the BFSS and BMN models). Given this observation it would interesting to identify the precise observables that survive this null limit, and to understand whether the deformation parameter appears as a Hamiltonian twist, a modified light-cone compactification, or an induced background flux.

Another noteworthy point is that the uni-vector deformations constructed in \cite{Gubarev:2025hvr} using the GL(D+1)-invariant formalism are not the only approach to deforming supergravity solutions with a uni-vector parameter. For instance, in \cite{Gubarev:2025qox}, such deformations were constructed for heterotic supergravity solutions using the O(d,d+N)-invariant formalism known as gauged double field theory. This raises several interesting questions regarding the relationship between these two types of uni-vector deformations: Are they equivalent? Can the O(d,d+N) deformations also be interpreted as diffeomorphisms in extended space? Do they also correspond to D0-branes or some other objects?

Finally, it would be interesting to investigate holographic interpretations of uni-vector deformations understood as adding dissolved D0-brane charge. For example, a natural question is whether uni-vector deformations admit a direct field-theory interpretation as a class of irrelevant deformations that inject particle-number, D0, or electric charge into the holographic state while preserving enough control over the bulk geometry. Given the relation between uni-vector deformations and the DLCQ limit, an obvious holographic direction is to understand the dual description in terms of discrete light-cone quantized gauge theories or matrix quantum mechanics. One would like to identify which observables are protected under the null limit, how the deformation parameter reorganizes the light-cone spectrum, and whether one can derive a holographic dictionary interpolating between a conventional brane dual and a DLCQ/matrix-model dual. This may also clarify the relation between uni-vector deformations and the BFSS/BMN sector of M-theory.

\section*{Acknowledgments}

The work of Kirill Gubarev was supported by the state assignment of the Institute for Information Transmission Problems of RAS.

\appendix

\section{D2-D0 made precise}
\label{sec:D2-D0-precise}

Let the smeared M2-brane extend along $(t,x^1,x^2)$, and let $z$ be the smeared transverse direction. Define
\begin{equation}
ds_7^2 \equiv dr^2+r^2 d\Omega_6^2,
\qquad
H(r)=1+\frac{Q}{r^5}.   
\end{equation}
After the rescaling $z \to \Lambda z,$ the 11d background becomes
\begin{equation}
\begin{aligned}
ds_{11}^2 & = H^{-2/3}\left(-dt^2+dx_1^2+dx_2^2\right) +
H^{1/3}\left(\Lambda^2 dz^2+ds_7^2\right), \\
A_3 &=(H^{-1}-1)\,dt\wedge dx^1\wedge dx^2.
\end{aligned}
\end{equation}

Now perform a boost with rapidity $\beta$:
\begin{equation}
\begin{aligned}
t_{\rm old}&=c\,t+s\,z, \\
z_{\rm old}&=s\,t+c\,z,
\qquad
\end{aligned}
\end{equation}
where we use $c=\cosh\beta$, $s=\sinh\beta$ for clarity of expressions. 
Then the metric and the 3-form can be written in KK form as
\begin{equation}
\begin{aligned}
ds_{11}^2
=&-\Lambda^2 H^{1/3}K_\Lambda^{-1}\,dt^2  +
H^{-2/3}(dx_1^2+dx_2^2)
+H^{1/3}ds_7^2
\\
&
+H^{-2/3}K_\Lambda
\left(
dz+\frac{sc(\Lambda^2 H-1)}{K_\Lambda}\,dt
\right)^2, \\
A_3  = & \
(H^{-1}-1)(c\,dt+s\,dz)\wedge dx^1\wedge dx^2,
\end{aligned}
\end{equation}
where we define 
\begin{equation}
K_\Lambda(H)\equiv \Lambda^2 c^2 H-s^2.    
\end{equation}

We now reduce the rescaled-boosted M2-brane background along $z$ using the  standard 11d/10d ansatz \eqref{eq:KK11to10}. The resulting type IIA background in the string frame is as follows
\begin{equation}
\label{eq:D2-D0resc}
\begin{aligned}
ds_{10,\text{str}}^2
&=
-\Lambda^2 K_\Lambda(H)^{-1/2}\,dt^2
+
H^{-1}K_\Lambda(H)^{1/2}(dx_1^2+dx_2^2) \\
&\qquad
+
K_\Lambda(H)^{1/2}(dr^2+r^2d\Omega_6^2),\\
e^\phi
&=
g_s\,H^{-1/2}K_\Lambda(H)^{3/4},\\
C_1
&=
\frac{\Lambda^2 sc}{\Lambda^2 c^2-s^2}\,
\frac{H-1}{K_\Lambda(H)}\,dt,\\
B_2
&=
s(H^{-1}-1)\,dx^1\wedge dx^2,\\
C_3
&=
c(H^{-1}-1)\,dt\wedge dx^1\wedge dx^2,
\end{aligned}
\end{equation}
with $K_\Lambda(H)=\Lambda^2 c^2 H-s^2$ and we choose the gauge were $C_1$ vanishes at infinity.

Certainly, for $\Lambda=1$, this reduces to the standard boosted-M2 $\to$ D2-D0 background with $K_\Lambda(H)\to c^2 H-s^2$. However, as we will see below, to compare to the result of a uni-vector deformation we have to keep the parameter $\Lambda$. What this parameter does is a controlled rescaling of the D0 charge on the D2 world-volume. Indeed, at large $r$ we have
\begin{equation}
C_3 \sim -c\,\frac{Q}{r^5}\,d\tau\wedge dx^1\wedge dx^2,
\qquad
C_1 \sim \frac{\Lambda^2 sc}{(\Lambda^2 c^2-s^2)^2}\,\frac{Q}{r^5}\,d\tau.
\end{equation}
For charges this implies $Q_{\text{D2}} \propto c\,Q$, $Q_{\text{D0}} \propto sc\,\L^2(\Lambda^2 c^2-s^2)^{-2}\,Q$, and therefore, we have two parameters to control  the ratio:
\begin{equation}
\frac{Q_{\text{D0}}}{Q_{\text{D2}}}\propto \frac{\Lambda^2 s}{(\Lambda^2 c^2-s^2)^2}.
\end{equation}
For the standard background $\L=1$ and the ratio is proportional to $\sinh \b$, while for our case, as we will see below, $\L = -\cosh \b /\sinh \b$. The ratio is still determined by the boost parameter. 
\begin{equation}
    \frac{Q_{\text{D0}}}{Q_{\text{D2}}}\propto \fr{c^2 s^3}{(c^2 + s^2)^2}
\end{equation}
and the naive singularity at $c^2 = s^2$ disappears.

Let us now compare this to the result of a uni-vector deformation. Since uni-vector deformations $t \to t + \a z$ considered here are parabolic transformations, in contrast to boosts, that are hyperbolic transformations, one cannot expect to arrive at the same results here. However, it appears, that although the 11d backgrounds are different, after the KK reduction and certain rescalings of coordinates one arrives at the same result. From the same pp-wave example we deduce that one should perform reduction along the light-like circle, therefor we turn to light-cone coordinates for the smeared M2-brane background
\begin{equation}
\tau=\frac{t+z}{\sqrt2},
\qquad
\zeta=\frac{t-z}{\sqrt2}.
\end{equation}
In these coordinates the background becomes
\begin{equation}
\begin{aligned}
ds_{11}^2 = &\ \frac{H^{1/3}-H^{-2/3}}{2}\left(d\tau^2+d\zeta^2\right)
-\left(H^{1/3}+H^{-2/3}\right)d\tau d\zeta
+H^{-2/3}(dx_1^2+dx_2^2)
+H^{1/3}ds_7^2 \\
A_3 = & \frac{H^{-1}-1}{\sqrt2}(d\tau+d\zeta)\wedge dx^1\wedge dx^2.
\end{aligned}
\end{equation}
Now perform rescaling of the coordinates $\t$, $\z$ and the shear transformation
\begin{equation}
\tau\to \nu \tau,
\qquad
\zeta\to \mu \zeta,
\qquad
\tau\to \tau+\alpha \zeta. 
\end{equation}
It is convenient to combine them as
$$
\hat\tau=\nu(\tau+\alpha\zeta),
\qquad
\hat\zeta=\mu\zeta,
$$
The metric and the 3-form become
\begin{equation}
\begin{aligned}
ds_{11}^2 = & H^{-2/3}K(H)\big(d\zeta+C_1\big)^2
-\frac{H^{1/3}\nu^2\mu^2}{K(H)}\,d\tau^2 +H^{-2/3}(dx_1^2+dx_2^2)
+H^{1/3}ds_7^2, \\
A_3 = &
\frac{H^{-1}-1}{\sqrt2}\Big[\nu\,d\tau+(\alpha\nu+\mu)\,d\zeta\Big]\wedge dx^1\wedge dx^2.
\end{aligned}
\end{equation}
where
\begin{equation}
    \begin{aligned}
        C_1= & \frac12 K(H)^{-1}\nu\big[H(\alpha\nu-\mu)-(\alpha\nu+\mu)\big]\,d\tau, \\
        K(H)= & \frac12\Big[H(\alpha\nu-\mu)^2-(\alpha\nu+\mu)^2\Big].
    \end{aligned}    
\end{equation}

We now reduce  along $\z$ using the standard 11d/10d ansatz \eqref{eq:KK11to10} and arrive at the following Type IIA solution
\begin{equation}
\label{eq:M2_deform}
\begin{aligned}
ds^2_{\text{str},\,\text{sh}}
= &
-\nu^2\mu^2\,K_{\text{sh}}(H)^{-1/2}\,d\tau^2
+
H^{-1}K_{\text{sh}}(H)^{1/2}(dx_1^2+dx_2^2)
+
K_{\text{sh}}(H)^{1/2}ds_7^2,\\
e^{\phi_{\text{sh}}}
= &
g_s^{(\text{sh})}\,H^{-1/2}K_{\text{sh}}(H)^{3/4},\\
C_{1,\text{sh}}= &
\nu\,\frac{Hu-v}{Hu^2-v^2}\,d\tau, \\
B_{2,\text{sh}}
= &
\frac{v}{\sqrt2}(H^{-1}-1)\,dx^1\wedge dx^2,\\
C_{3,\text{sh}}
= &
\frac{\nu}{\sqrt2}(H^{-1}-1)\,d\tau\wedge dx^1\wedge dx^2,\\
K_{\text{sh}}(H)&=\frac12\big(Hu^2-v^2\big).
\end{aligned}
\end{equation}
with
$$
u\equiv \alpha\nu-\mu,\qquad v\equiv \alpha\nu+\mu.
$$

In the gauge vanishing at infinity,
$$
C_{1,\text{sh}}^{(\infty=0)}
=
\nu\,\frac{uv}{u+v}\,\frac{H-1}{Hu^2-v^2}\,d\tau.
$$

Let us now show that the background \eqref{eq:M2_deform} is nothing but \eqref{eq:D2-D0resc}. Start with matching the metrics. Because the transverse piece $ds_7^2$ is untouched, we must match the harmonic prefactor exactly. So we need $K_{\text{sh}}(H)=K_\Lambda(H)$. This implies 
\begin{equation}
u^2= 2\Lambda^2 c^2,\qquad v^2=2 s^2.
\end{equation}
To match $B_2$ with the same sign as in the boosted solution, choose
$v=\sqrt2\,s$. To match the graviphoton field $C_1$ in the asymptotically vanishing gauge, the correct sign choice is
$u=-\sqrt2\,\Lambda c$. Therefore we have
$$
\alpha\nu-\mu=-\sqrt2\,\Lambda c,\qquad
\alpha\nu+\mu=\sqrt2\,s.
$$
This gives 
$$
K_{\text{sh}}(H)=K_\Lambda(H)=\Lambda^2 c^2 H-s^2.
$$
As in the case of generation of bound states by poly-vector deformations \cite{Barakin:2025jwp}, one has to rescale time. The correct rescaling appears to be
$$
t=\frac{\nu\mu}{\Lambda}\,\tau,
$$
so the string-frame metrics match exactly:
$$
ds^2_{\text{str},\,\text{sh}}=ds^2_{\text{str},\,\Lambda}.
$$
The dilaton fields in both solutions match if we simply identify $
g_s^{(\text{sh})}=g_s^{(\Lambda)}$. Therefore, no non-trivial redefinition of $g_s$ is actually needed once $K$ is matched exactly.

For the NS-NS field $B_2$ using $v=\sqrt2 s$ we obtain
$$
B_{2,\text{sh}}
=
\frac{v}{\sqrt2}(H^{-1}-1)\,dx^1\wedge dx^2
=
s(H^{-1}-1)\,dx^1\wedge dx^2=B_{2,\Lambda}.
$$

Similarly, for the  RR 1-form $C_1$ in the gauge with $C_1\to 0$ at infinity, we have
$$
C_{1,\text{sh}}^{(\infty=0)}
=
\nu\,\frac{uv}{u+v}\,\frac{H-1}{Hu^2-v^2}\,d\tau = \frac{\Lambda^2 sc}{\Lambda^2 c^2-s^2}\,
\frac{H-1}{K_\Lambda(H)}\,dt = C_{1,\Lambda}^{(\infty=0)}.
$$
Here we substitute
$$
u=-\sqrt2\Lambda c,\qquad v=\sqrt2 s,\qquad
u+v=\sqrt2(s-\Lambda c),
$$
and then convert to $t$ using $t=\nu\mu \Lambda^{-1}\tau$, with
$$
\mu=\frac{s+\Lambda c}{\sqrt2}.
$$

Finally, after the time rescaling we have the following for the R-R 3-form field
$$
C_{3,\text{sh}}
=
\frac{\nu}{\sqrt2}(H^{-1}-1)\,d\tau\wedge dx^1\wedge dx^2
=
\frac{\Lambda}{s+\Lambda c}\,
(H^{-1}-1)\,dt\wedge dx^1\wedge dx^2.
$$
To match the boosted result $
C_{3,\Lambda}
=
c(H^{-1}-1)\,dt\wedge dx^1\wedge dx^2.
$
we set $c = - \L s$, that is the precise reason for the introduction of the additional rescaling. 

\bibliographystyle{utphys}
\bibliography{bib.bib}

\end{document}